\documentclass[
bibnotes,
 amsmath,amssymb,
 aps]{revtex4}
\usepackage{amssymb, amsthm}
\usepackage{txfonts, comment}
\usepackage{mathtools}
\usepackage{bm}
\usepackage[dvipdfm]{graphicx}
\usepackage{ascmac}
\usepackage{booktabs}
\usepackage{wrapfig}
\usepackage{tikz, ifdraft}
\usepackage{physics}
\usepackage{verbatim}
\usepackage{algorithm}
\usepackage{algpseudocode}
\usepackage{blindtext}
\usepackage{subfigure}
\usepackage{times}
\usepackage{txfonts}
\usepackage{siunitx}
\usepackage{array}
\usepackage[driverfallback=dvipdfm]{hyperref}
\newcolumntype{C}[1]{>{\centering\let\newline\\\arraybackslash\hspace{0pt}}m{#1}}
\newcolumntype{R}[1]{>{\raggedleft\let\newline\\\arraybackslash\hspace{0pt}}m{#1}}
\newcommand{\argmax}{\mathop{\rm arg~max}\limits}
\newcommand{\argmin}{\mathop{\rm arg~min}\limits}

\newcommand{\E}{\mathbb{E}}

\newcommand{\Kmax}{{K_{\rm max}}}\newcommand{\Kmin}{{K_{\rm min}}}
\newcommand{\THeta}{{\vb*{\theta}}}

\newcommand{\model}{{\mathcal{M}}}
\begin{document}
\title{Fast Bayesian Deconvolution using Simple Reversible Jump Moves}
\author{Koki Okajima$^1$, Kenji Nagata$^2$, and Masato Okada$^{2,3}$\footnote{okada@edu.k.u-tokyo.ac.jp}}
\affiliation{$^1$Graduate School of Science, The University of Tokyo, Bunkyo, Tokyo 113-0033, Japan  \\
$^2$Research and Services Division of Materials Data and Integrated System, National Institute for Materials Science, Tsukuba, Ibaraki 305-0047, Japan \\
$^3$Graduate School of Frontier Science, The University of Tokyo, Kashiwa, Chiba 277-8561, Japan} 
\begin{abstract}We propose a Markov chain Monte Carlo-based deconvolution method designed to estimate the number of peaks in spectral data, along with 
      the optimal parameters of each radial basis function. 
      Assuming cases where the number of peaks is unknown, and a sweep simulation on all candidate models is computationally unrealistic, the proposed method efficiently searches over the probable candidates via trans-dimensional moves assisted by annealing effects from replica exchange Monte Carlo moves. Through simulation using synthetic data, the proposed method demonstrates its advantages over conventional sweep simulations, particularly in model selection problems. Application to a set of olivine reflectance spectral data with varying forsterite and fayalite mixture ratios reproduced results obtained from previous mineralogical research, 
      indicating that our method is applicable to deconvolution on real data sets.
\end{abstract}
\maketitle
\section{Introduction}
Spectral analysis is a primary tool for examining matter. 
Information on the peak location and intensity within the reflectance spectra is valuable for
specifying mineralogical properties, such as the composition and crystal structure\cite{sunshine1990}. 
However, deconvoluting noisy spectral data with overlapping peaks is generally a difficult task, as it involves 
regression using nonlinear functions. Simple methods, such as least squares regression using gradient methods often fail owing to  
local minima in the parameter space\cite{Nagata2012,ikeda2015}. Moreover, since the number of peaks is often unknown, 
it is necessary to determine the number of basis functions to be used for regression in this process. \\
\indent The task of deconvolution is not a simple regression problem. 
It is a problem of inferring the hidden structure within the data: in this case, the number 
of peaks and the width, 
position, and intensity of each peak. 
To do this, an appropriate model must be assumed for this structure. This is performed by model selection, 
where the true model is inferred from a set of candidate models\cite{Bishop}. In model selection, it is appropriate to compare 
the Bayesian free energy of each model\cite{Nagata2012,Akaike1980} against other model selection criteria,
 such as the widely applicable information criterion (WAIC)\cite{watanabe2010} or 
cross-validation evaluators.
\\
\indent The calculation of the Bayesian free energy requires a high-dimensional numerical integration, 
which is performed practically by Markov chain Monte Carlo (MCMC) methods. 
The reversible jump MCMC (RJMCMC) method is of the sampling type originally designed for model selection, 
which searches over a union set of candidate models and annexed parameters\cite{green1995, Razul2001, Han2020}, and calculates the posterior model probability, informatively equivalent to the Bayesian free energy of each model.
On the other hand, Nagata et al. have proposed a deconvolution method for calculating the Bayesian free energy of a single model\cite{Nagata2012}. 
This is another MCMC sampling method, but with replica exchange
to achieve the global minimum of the regression problem\cite{hukushima1996}. 
Even though model selection is performed by a sweep over the candidate model set,
 the method is reported to be able to determine the optimal model and regression parameters more rapidly and more accurately 
 than the RJMCMC method with the 
replica exchange method combined, a technique previously exploited for the mixture model by Jasra et al. \cite{jasra2005}. \\ %
\indent In this study, which is in contrast to previous studies, we propose a MCMC method sampling over the union set of candidate models and its parameters.
This method consists of a simple pair of RJMCMC moves and the replica exchange move. 
 Although the proposed method requires a search over a much more complex energy landscape, 
the global minimum and the posterior model probabilities were successfully obtained 
with less computation time than the conventional method. 
This was confirmed from numerical experiments conducted on multipeak synthetic data by closely examining the convergence time in terms of the burnin and sampling period, as well as the accuracy of model selection under limited computational resources.\\
\indent To further demonstrate its strength in real data applications, we apply this method to a 
set of reflectance spectral data of olivine with different mixture ratios of forsterite and fayalite, and show that the results support those obtained in mineralogical research.
\\
\indent The outline of this paper is as follows. In sect. 2, we describe the deconvolution model and its Bayesian framework of model selection. 
Although the model is described as a sum of Gaussians, we stress that it can be naturally applied to any other basis function. 
 In Sect. 3, the proposed MCMC method is described in detail, whereas the setup and results of the numerical experiment comparing the proposed and conventional methods are given in Sect. 4. Finally, the real data analysis of the reflectance spectral data of olivine is given in Sect. 5.
\section{Problem and Formulation}
\subsection{Model}
Suppose our model assumes that the observation $y\in \mathbb{R}$ is given by the sum of $K$ Gaussian peaks distorted by Gaussian noise:
\begin{equation}\label{data}
y = \sum_{k = 1}^Kf (x;\theta_k) + \epsilon,\ \epsilon \sim \mathcal{N}(0, 1/b) ,
\end{equation}
\begin{equation}
f(x;\theta_k) = a_k \exp\qty[ -\frac{\rho_k}{2}(x-\mu_k)^2 ],
\end{equation}
where the position of each observation $x\in\mathbb{R}$ is given, $b$ is the inverse of the noise variance, namely, the observation precision, and 
$\Theta_K = \{\theta_k\}_{k = 1}^K, \ \theta_k = \{a_k, \rho_k, \mu_k\}$ are parameters. Moreover, $a_k, \rho_k,$ and $\mu_k$ 
are the intensity, precision, and center of each peak, respectively. 
We also define the function $f(x; \theta_0 = \emptyset) = 0$ for $K = 0$, which corresponds to the case 
where no Gaussian peaks are assumed in the model. 
Although the basis function $f$ is taken to be Gaussian, it can be set to an arbitrary
 function without loss of generality.\\
 \indent Under this assumption, the observations are subject to the following conditional probability:
 \begin{equation}
p(y|x,\Theta_K, K, b) = \qty(\frac{b}{2\pi})^{1/2}\exp\qty{-\frac{b}{2} \qty[ y-\sum_{k = 1}^K f(x; \theta_k) ]^2}. 
 \end{equation}
For a set of independent observations $\mathcal{D} = \{(x_i, y_i)\}_{i = 1}^n$, the conditional probability is given by \begin{align}
p(\mathcal{D} |\Theta_K,K, b) &= \qty(\frac{b}{2\pi})^{n/2} \exp\qty[ -nbE_n(\Theta_K) ]  ,      
\end{align}
where the energy $E_n$ is defined
as follows by the mean squared
error of the observation $\mathcal{D}$ and the regression function: \begin{equation}
E_n(\Theta_K) = \frac{1}{2n}\sum_{i = 1}^n \qty[ y_i-\sum_{k = 1}^K f(x_i; \theta_k) ]^2.  
\end{equation}
\subsection{Bayesian Framework}
In the Bayesian framework, the parameters $\Theta_K$ are inferred from the posterior probability density given by Bayes' theorem:
\begin{equation}\label{eq:conv_post}
p(\Theta_K | \mathcal{D}, K, b) = \frac{1}{Z_n(K,b)} \qty(\frac{b}{2\pi})^{n/2} \exp\qty[-nbE_n(\Theta_K)]p(\Theta_K) ,\\
\end{equation}
\begin{equation}\label{eq:cond_partition}
Z_n(K, b)  = \qty(\frac{b}{2\pi})^{n/2} \int \qty(\prod_{k = 1}^K \dd \theta_k)\exp\qty[-nbE_n(\Theta_K)]p(\Theta_K),
\end{equation}
where $p(\Theta_K)$ is the prior probability of the parameters. 
The function $Z_n(K,b)$ is the normalization term called the partition function, and its 
negative logarithm is 
defined as the Bayesian free energy:\begin{equation}\label{eq:free}
F_n(K,b) = -\log Z_n(K,b).         
\end{equation}
Practically, the number of peaks $K$ and precision $b$ are often unknown. 
In previous works, these variables were estimated using the empirical Bayesian approach, where 
their estimators $\hat{K}$ and $\hat{b}$ were given by the maximizer of the Bayes free energy\cite{tokuda2017}.
 The number of peaks $K$ can also be treated as a random 
variable equally to the parameters $\Theta_K$. In fact, a posterior probability on the union set $\model = \cup_{K = \Kmin}^{\Kmax}\{K, \Theta_K\}$ can be investigated. Assuming a prior probability on the natural number $K\in \qty[\Kmin, \Kmax]$ as $p(K)$, this posterior distribution is given once again by Bayes' theorem:
\begin{equation}\label{eq:target}
          p(\Theta_K, K | \mathcal{D}, b)  =  \frac{1}{Z_n(b)} \qty(\frac{b}{2\pi})^{n/2}\ p(K) p(\Theta_K) \exp[-nb E_n(\Theta_K)],
\end{equation}
\begin{align}\label{eq:partition}
      Z_n(b) &= \qty(\frac{b}{2\pi})^{n/2}  \sum_{K = \Kmin}^{\Kmax}p(K) \int \qty(\prod_{k = 1}^K \dd \theta_k ) p(\Theta_K) \exp\qty[-nbE_n(\Theta_K)]\\
      \label{eq:dd}
      &= \sum_{K = \Kmin}^\Kmax p(K)Z_n(K, b).
\end{align}
\indent Correspondingly, the Bayes free energy is defined by\begin{equation}
      F_n(b) = -\log Z_n(b).        
      \end{equation} 
      Although we can estimate the noise in the Bayesian framework by designing a prior distribution for $b$, $b$ is assumed to be evaluated using the empirical Bayesian approach: \begin{equation}\label{eq:empirical}
            \hat{b} = \argmin_b F_n(b),
      \end{equation}
      where $\hat{b}$ is the estimator of the precision.
\subsection{Posterior Model Probability}
The posterior model probability is obtained by marginalizing Eq. \eqref{eq:target}:
\begin{equation}\label{eq:postK_proposed}
      p(\tilde{K} | \mathcal{D}, b) = \E_{\Theta_K, K |\ \cdot \ }\qty[\delta(K, \tilde{K})],
\end{equation}
where $\E_{\Theta_K, K |\ \cdot \ }\qty[\cdot]$ denotes the expectation over the probability density $p(K, \Theta_K | \mathcal{D}, b)$, and $\delta(K, \tilde{K})$ is the classic delta function. 
The same value can be obtained as follows if $Z_n(K,b)$ is known from Eq. \eqref{eq:dd}:
\begin{equation}\label{eq:postK_conventional}
      p(\tilde{K} | \mathcal{D}, b) = \frac{p(\tilde{K})Z_n(\tilde{K}, b)}{\sum_{K^\prime = \Kmin}^\Kmax p(K^\prime) Z_n(K^\prime, b)}.
\end{equation} 
The partition function $Z_n(K,b)$ can be calculated by sampling from the posterior density 
$p(\Theta_K| \mathcal{D}, K, b)$ in the range $[\Kmin, \Kmax]$, as in the conventional method\cite{Nagata2012,tokuda2017}.
This sequence of sampling can be computationally intensive, particularly in situations where 
$\Kmax-\Kmin$ is large, i.e., $K$ is very ambiguous, or the data simply resemble a large number of peaks. 
On the other hand,
 a significant speedup can be expected by using Eq. \eqref{eq:postK_proposed}, 
 as it only requires a single sampling task in the union set $\model$. 
 The main objective of this study is to design an appropriate MCMC sampling scheme 
for this union set and demonstrate the speedup achieved by employing this technique. 
\section{Proposed Method}
\subsection{Exchange Monte Carlo (EMC) Method}
Sampling from the posterior density $p(\Theta_K | \mathcal{D}, K, b)$ is prone to local minima in the energy landscape\cite{Nagata2012,tokuda2017}. 
Since the target density $p(\Theta_K, K  | \mathcal{D}, b)$ additionally incorporates discrete model variables,
it is obvious that the effect of local minima is more severe than that of the previous density.
 To overcome this, the EMC method is employed. 
 The target density of this MCMC scheme is the joint of several posterior densities with different observation precisions, namely, inverse temperatures, given by$ \prod_{l = 1}^L p(\Theta_{K^{(l)}}, K^{(l)} | \mathcal{D}, b_l)$, where $K^{(l)}$ denotes the model variable in the replica with the inverse temperature $b_l$.
Given a sequence of inverse temperatures $0 = b_1 < b_2 < \cdots < b_L = b$, the exchange of states enables low-temperature states trapped at local minima to 
reach the global minimum via relaxation in higher temperature regimes\cite{hukushima1996}. 
The above definition of the inverse temperature is according to that of Tokuda et al.\cite{tokuda2017}, which is different from
the more common notation in singular learning theory adopted by Watanabe \cite{watanabe2010} and Nagata et al.\cite{Nagata2012}. 
Not only does the EMC method prevent the freezing of states at local minima, 
it also allows the calculation of the partition function $Z_n(b)$ in a single run. Consider the non-normalized 
partition function $z_n(b)$ defined as \begin{equation}
z_n(b) =  \sum_{K = \Kmin}^\Kmax p(K) \int \dd \Theta_K p(\Theta_K) \exp \qty[ -nbE_n(\Theta_K) ].
\end{equation}
From bridge sampling\cite{ogata1990, gelman1998}, $z_n(b)$ is calculated by \begin{equation}
      \label{eq:calc_z}
z_n(b) = \prod_{l = 1}^{L-1} \ \E_{\Theta_K, K  | \ \cdot}^{b_l} \qty[\exp \qty[-n(b_{l +1}- b_l) E_n(\Theta_K)] ].
\end{equation}
The free energy at each replica inverse temperature $b_l$ can then be obtained from Eqs. \eqref{eq:partition} and \eqref{eq:calc_z}: 
\begin{equation}
      F_n(b_l) = -\frac{n}{2}\log \frac{b_l}{2\pi} - \log z_n(b_l).      
\end{equation}
By setting the inverse temperatures at each replica appropriately such that replica exchange is not impeded by a large temperature difference\cite{tokuda2014,Katzgraber2019},
we can obtain simultaneously the free energy $F_n(b_l)$ and posterior density $p(w | \mathcal{D}, b_l)$ at 
each replica temperature. \\
\indent Since the free energy is obtained at discrete replica temperatures, 
the noise variance cannot be straightforwardly estimated using Eq. \eqref{eq:empirical}. 
We will only mention that this can be performed by interpolating the free energy and 
posterior model probability using the multihistogram method\cite{Swensden1989}, which was carried out in the 
previous work by Tokuda et al.\cite{tokuda2017}.
\subsection{Reversible Jump Moves}
The RJMCMC method is a Metropolis-Hastings-based trans-dimensional MCMC move. As 
the name suggests, the method consists of a reversible pair of moves that alters the number of parameters; hence, it is a jump move from one model to another. Given the current model $\{\Theta_K \in \mathbb{R}^{N}, K\}$, a model with higher-dimension $\{\Theta_{K + 1}\in \mathbb{R}^{N + M}, K + 1\}$ is proposed via a deterministic function $h : \mathbb{R}^N \times \mathbb{R}^M \rightarrow \mathbb{R}^{N + M},\ 
(\Theta_K, \vb{u}) \mapsto h(\Theta_K, \vb{u}) = \Theta_{K + 1},$ and a random vector $\vb{u}\in \mathbb{R}^M$ generated from the distribution $g(\vb{u})$. For this proposal, the acceptance probability $\mathcal{A}$ is given by \begin{equation}
\mathcal{A} = \min\qty[ 1, \frac{p(\Theta_{K + 1}, K+1 | \cdot) r_{m^\prime}( K + 1)}{p(\Theta_K, K | \cdot)r_{m}( K) g(\vb{u})} \qty|\pdv{\Theta_{K + 1}}{(\Theta_K, \vb{u})} | ],
\end{equation}
where $r_m$ and $r_{m^\prime}$ are the probabilities of proposing the move above and the reversible move.
The acceptance probability of the reverse move is naturally defined by the inverse of $\mathcal{A}$. \\
\indent For our particular setup, we assume that the prior densities of each peak are independent of one another, i.e., $p(\Theta_{K + 1}) / p(\Theta_K) = p(\theta_{K + 1})$. Choosing $h$ to be an identity map and the random vector probability density to be equal to the prior density of the new parameters, i.e., $g(\vb{u}) = p(\theta_{K + 1})$, the acceptance probability is reduced to 
\begin{equation}
    \mathcal{A} = \min \qty[ 1, \frac{p(\mathcal{D} | \Theta_{K + 1}, K + 1, b)r_{m^\prime}(K + 1) }{p(\mathcal{D} | \Theta_K, K ,b)r_{m}(K)} ].
\end{equation}
Finally, we define the proposal probabilities as $r_m(K) = 1/2 \ (K = \Kmin + 1, \cdots , \Kmax -1)$,  $r_m(\Kmin) = 1, r_m(\Kmax) = 0$, and $r_m(K) + r_{m^\prime}(K) = 1$.\\
\indent Reversible jumps in low temperature regimes are rarely accepted in general, because such moves markedly deform the regression function. 
The proposed method thus relies on the mixing properties of the replica exchange moves for trans–dimensional updates at lower temperatures. 
We argue that more sophisticated reversible jump moves, such as the peak split/merge moves\cite{Razul2001, Han2020} do not improve the algorithm, but rather decrease its generality, as such moves must be tailored specifically for each application to keep the moves reversible. 
\subsection{Local Updates and Step-Size Tuning}
\indent Local updates are performed on each parameter according to the standard Metropolis–Hastings algorithm at each replica in parallel to ensure ergodicity. 
For adequate sampling at each replica, the step size for $\theta$ was designed appropriately.
 For instance, at high inverse temperatures, the target distribution $p(\Theta_K, K | \mathcal{D}, b)$ is strongly peaked at the 
minima of the energy $E_n(\Theta_K)$. 
\begin{algorithm}[H]
      \caption{Adaptive step size tuning}
            \label{alg:tuning}
      \begin{algorithmic}
            \State \textbf{Parameters:} $c_t, t_0, p^\star$ 
            \State \textbf{Require:} $p_{\rm accept}$, acceptance rate of MC step.
            \State \textbf{Initialize:} $\Delta\THeta$, initial step size; $M_{\rm max}$, number of samples to calculate $p_{\rm accept}$; $t_{\rm max}$, total number of updates. 
             \For{$t = 0, 1, \cdots t_{\rm max}$}
                    \State $p_{\rm accept} = 1.0$
                    \For{$M = 0,1, \cdots M_{\rm max}$}
                           \State MC state using step size $\Delta \THeta_{t}$ 
                           \If{accepted}
                                  \State{$p_{\rm accept} \leftarrow (p_{\rm accept} M + 1)/(M + 1)$}
                           \Else
                                  \State{$p_{\rm accept} \leftarrow p_{\rm accept}  M /(M + 1)$}
                           \EndIf
                    \EndFor
                    \State \textbf{Update step size:} $\Delta\THeta\leftarrow \Delta\THeta + c_t(p-p^\star)/(t_0 + t)$
             \EndFor
      \end{algorithmic}
  \end{algorithm}
\begin{figure*}[b]
      \centering 
      \subfigure[]{
      \includegraphics{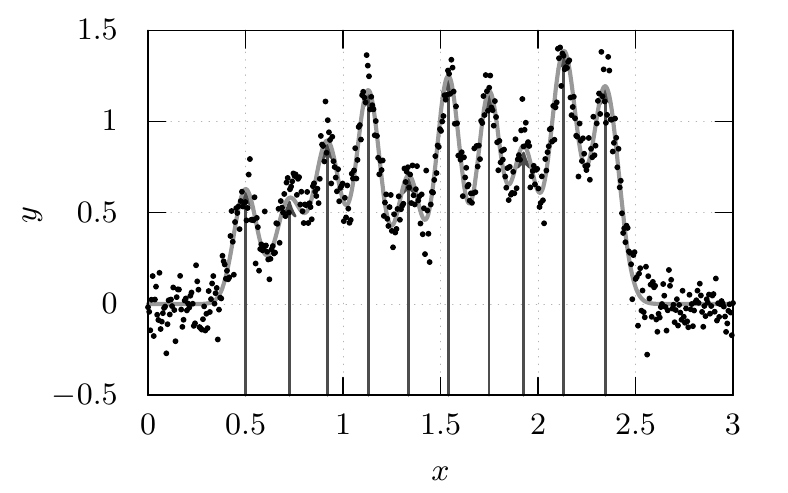}%
      }%
      \subfigure[]{%
            \includegraphics{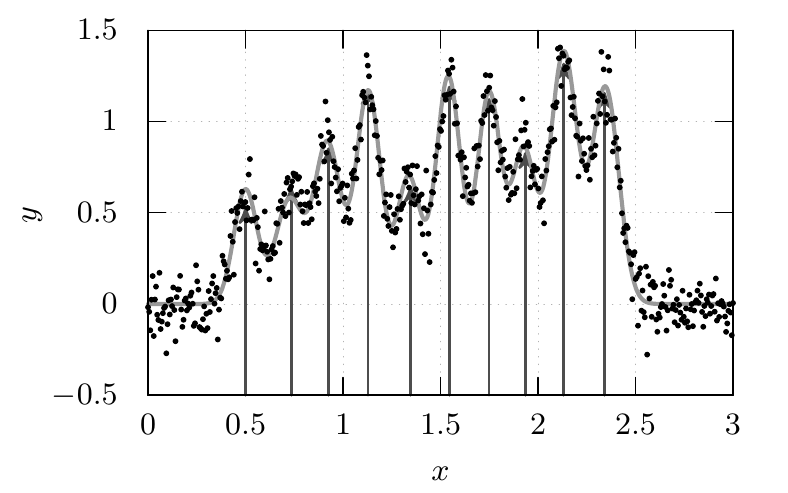}%
      }%
      \caption{(a) Synthetic data with 10 peaks used in the simulation. (b) Deconvolution results obtained using the proposed method at inverse temperature $b = 100$. 
      In (a), the vertical arrows represent the position and height of each peak used in generating the data. In (b), the vertical arrows represent the 
      position and height of each convoluted peak. The position and height of each convoluted peak agree well with the true values.}
      \label{fig:data}
\end{figure*}
\begin{figure*}
      \centering
      \includegraphics{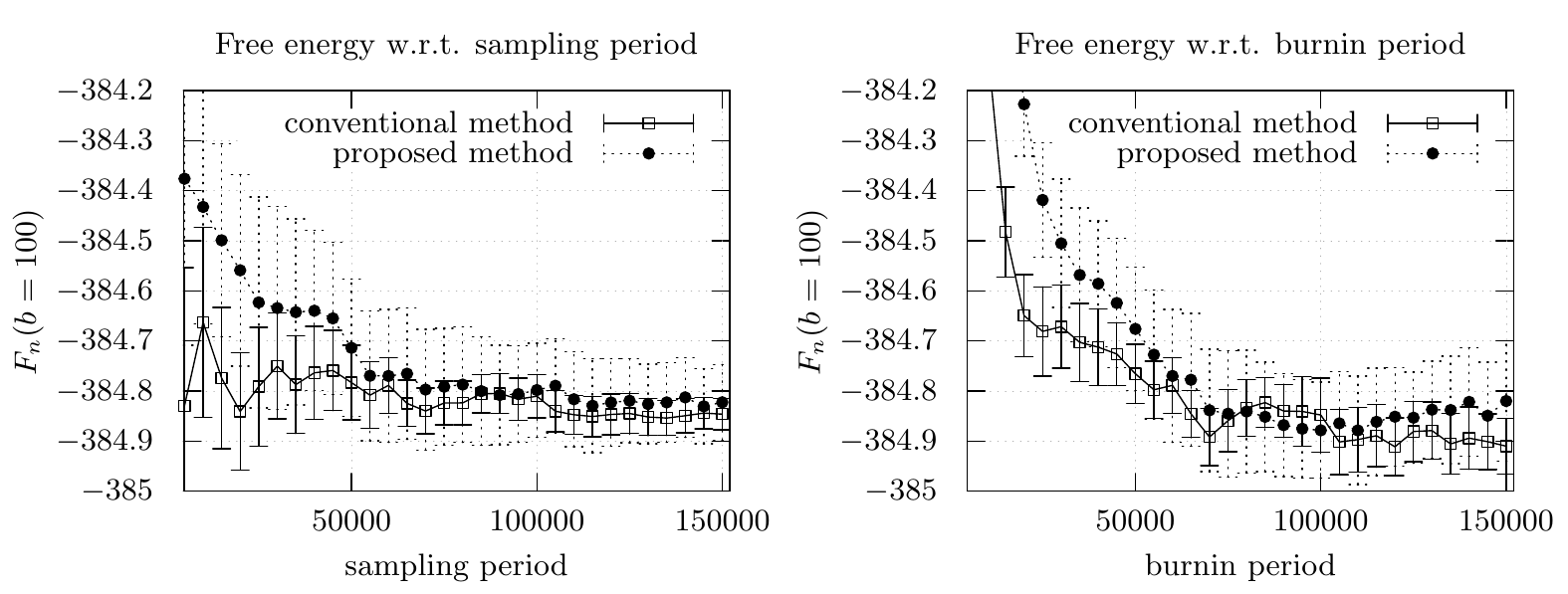}
      \caption{Free energy with respect to different sampling periods (left) under 60,000 burnin periods and different burnin periods (right) under 60,000 sampling periods. 
       Note that the figures are complementary; the left figure indicates that 60,000 MCSs are enough for the sampling period, whereas the right figure 
       indicates that 60,000 MCSs are sufficient for the burnin period. Error bars are given by the standard error for 20 independent runs. }
      \label{fig:FE_dynamics}
  \end{figure*}
One can imagine that a large step size for the continuous parameters will be strongly rejected. 
In fact, the step size should not be universal throughout all replicas, but it should decrease with the replica temperature. \\
\indent Here, an adaptive step-size tuning method proposed by Garthwaite et al.\cite{garthwaite2016}, which keeps the acceptance rate at a predetermined constant value, is adopted. 
This is carried out by enforcing the Robbins–Monro algorithm\cite{robbins1951} and updating the step size $\Delta\theta(b)$ for 
the replica inverse temperature $b$ as \begin{equation}
      \Delta\theta_{t + 1}(b) = \Delta \theta_t(b) + c_t\frac{ p(b)-p^\star}{t + t_0},
\end{equation}
where $t,p^\star$, and $p(b)$ denote the iterative step for the update, the target acceptance probability, and the current acceptance probability at
the replica with inverse temperature $b$, respectively. 
The parameters $c_t$ and $t_0$ are universal in all replicas and are determined heuristically. The pseudo-algorithm is given in Algorithm \refeq{alg:tuning}.
Although heuristics are involved, only two parameters need to be tuned for optimal performance, making this a relatively simple task. \\
\indent The proposed method consists of the three Monte Carlo moves given above. A single Monte Carlo step (MCS) is 
defined by a reversible jump move and a local update move on all replicas done alternately ten times consecutively, followed by a single EMC move. 

\section{Simulation using Synthetic Data}
\begin{figure*}[t]
      \centering
             \includegraphics{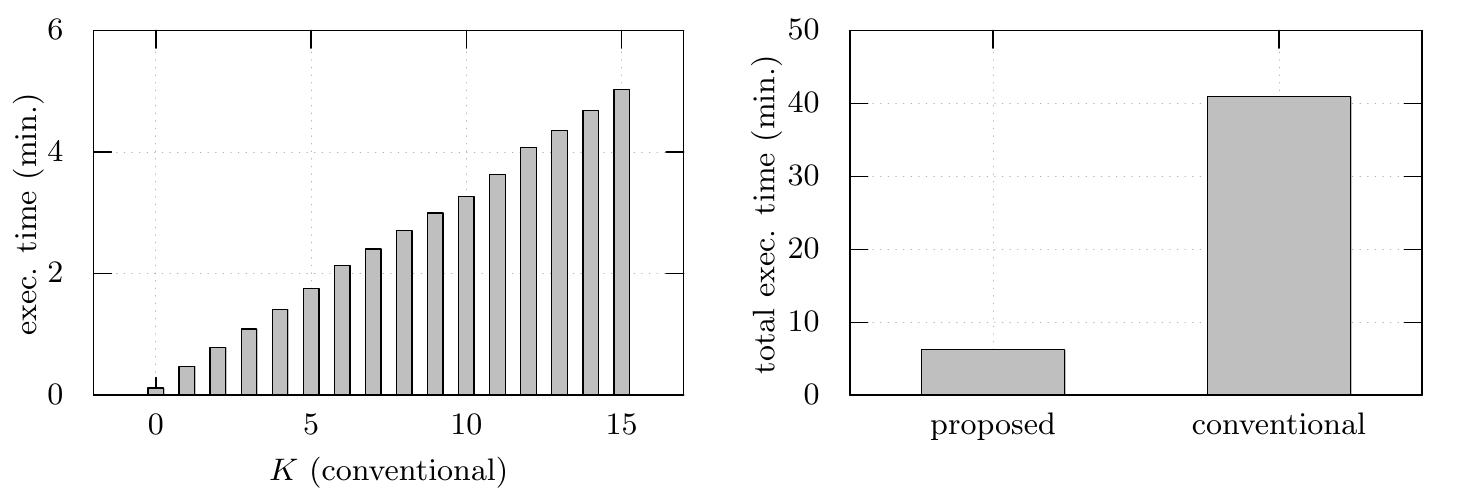}
             \caption{Execution time of the conventional method for each model with $K$ peaks (left panel) and the total execution time 
             for each method (right). In the conventional method, the execution time scales with $K$, indicating a total computation time of $O(\Kmax^2/2)$ 
             for the entire model selection process. 
             All numerical experiments were performed using an AMD Ryzen(TM) 3700X CPU, which has 8 cores (16 threads) at 4.0 GHz with 32 GB memory.}
      \label{fig:time}
\end{figure*}

Synthetic data $\mathcal{D}$ were generated according to Eq. \eqref{data}, with the true number of peaks $K_0 = 10$, 
inverse noise variance $b_0 = 100$, and $n = 512$ with $x$ distributed equidistantly at the interval $[0,3]$.
 The actual generated data are given in Fig. \ref{fig:data}(a). 
The prior density is defined by \begin{equation}
p(\Theta_K) = \prod_{k = 1}^K \varphi(\theta_k), 
\end{equation}
\begin{equation}
\varphi(\theta_k) = {\rm Gam} (a_k; 5, 5)\ {\rm Gam}(b_k, 5, 0.04)\ \mathcal{U}(\mu_k;\qty[0,3]),   
\end{equation}
where ${\rm Gam}$ is the Gamma distribution defined by \begin{equation}
{\rm Gam}(x; \eta ,\lambda) = \frac{\lambda^\eta x^{\eta -1}}{\Gamma(\eta)} \exp(-\lambda x),      
\end{equation}
and $\mathcal{U}(x; I)$ is the uniform distribution defined at interval $I$.  
The prior distribution for the number of peaks was set to a uniform distribution, assuming that we have no information about the 
number of peaks. 
For the EMC method, the number of replicas was set to $L = 60$, with the replica inverse temperatures given by \begin{equation}
b_l = \begin{cases}
      0 & (l = 1)\\
      144 \cdot 1.2^{l-L} & (\text{otherwise})
\end{cases}.
\end{equation}
In practice, the maximum replica inverse temperature $b_L$ should be set higher than 
$\hat{b}$ to obtain the minimum free energy using the multihistogram method\cite{tokuda2017}. 
Here, we used $b_L = 144$, considerably higher than the ground-truth value $b_0 = 100$. 
The minimum and maximum numbers of peaks were set to $\Kmin= 0$, and $\Kmax = 15$, repectively.\\
\indent To validate our results, simulations on independent runs for $K$\cite{Nagata2012,tokuda2017}
 with the same step-size tuning method were also performed. Although this conventional method only has continuous parameters, 
our method also includes the model variables. It is important to confirm that whether the proposed method converges as rapidly
 as the conventional method even with trans-dimensional moves. 
 \begin{figure*}[t]
      \centering
      \includegraphics{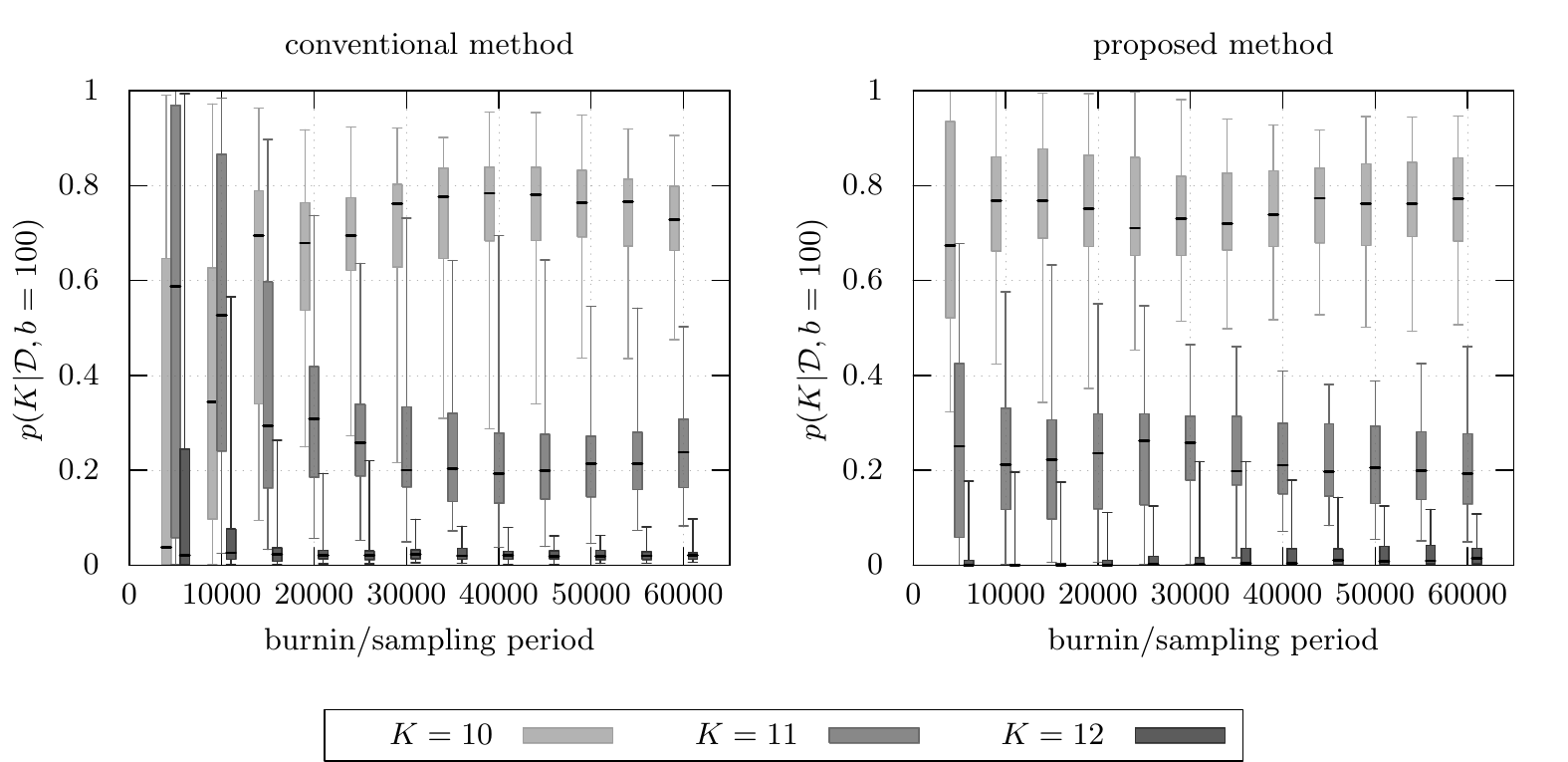}
      \caption{Posterior probability of the number of peaks for the conventional and proposed methods with the same burnin and sampling periods, averaged over 40 independent runs.
       The conventional method indicates a high model misselection rate ($>25\ \%$) under 40,000 MCSs for burnin and sampling periods, whereas the proposed method is highly accurate even with 10,000 MCSs for burnin and sampling periods.}
      \label{fig:boxplot}
  \end{figure*}
  \begin{figure*}
      \centering
      \includegraphics{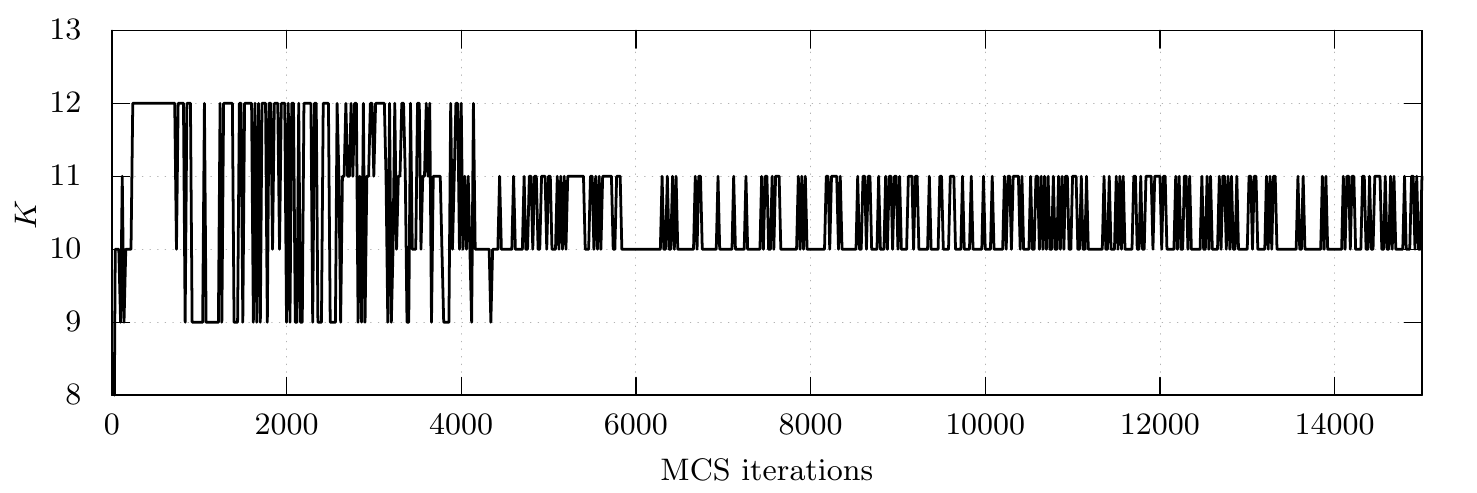}
      \caption{Mixing property of $K$ for the proposed method. The values of $K$ for every 20 MCS are plotted. The MCMC sampler $K$ intensively searches model $K = 10$ and $11$ when iterations exceed 5,000 MCSs, which is in agreement with Fig. \ref{fig:boxplot}.}
      \label{fig:mix}
  \end{figure*}
 Using both methods, we ran 20 independent simulations on the data given in Fig. \ref{fig:data} for 210,000 MCSs in total to investigate the burnin and sampling period necessary for sufficient sampling. 
 Also, 40 independent simulations for relatively short MCSs (maximum of 120,000 MCSs) were carried out to compare performance, assuming situations where computation resources are limited.\\
 \indent Figure \ref{fig:FE_dynamics} shows the free energies obtained using the conventional and proposed methods with different burnin and sampling periods. 
 As evident from the graph, both methods need approximately 60,000 MCSs for thermalization to obtain reliable free energy values under the fixed sampling period of 60,000 MCSs. 
 Under the assumption that 60,000 MCSs are sufficient for thermalization for both methods, performance differences are evident with respect to the sampling period. 
The proposed method requires approximately 100,000 MCSs to obtain the free energy a standard error of less than $0.1$,
whereas the conventional method can acquire the same quality of value in less than 50,000 MCSs.
 In terms of deviation from the true free energy value, estimated to be about $-384.83$, the proposed method requires a sampling period of 60,000 MCSs 
 to obtain a nonbiased value, whereas the sampling period necessary for the conventional method is minimal. 
 These results do not necessarily indicate that the conventional method overperforms the proposed method.
 Note that the specified sampling and burnin periods is taken on all candidate models in the proposed method;
  therefore, the substantial number of MCSs for the conventional method is $\Kmax- \Kmin$-fold. \\
 \indent Figure \ref{fig:boxplot} shows the posterior probability of the number of peaks for $K = 10,11,12$. 
 The probabilites of other number of peaks were omitted since these values were irrelevant. 
 Clearly, the proposed method shows a lower model mis-selection rate under burnin and sampling periods of less than 60,000 MCSs than the conventional method. 
 In fact, the proposed method had zero misselection for sampling and burnin periods above 30,000 MCSs. 
 The performance of the model mixing of the proposed method can also be verified from the model mixing property indicated in Fig. \ref{fig:mix}.  \\
\indent Figure \ref{fig:time} shows the computation time necessary for a single run in each method under equal sampling and burnin periods. 
Since the conventional method can only obtain the free energy $F_n(b)$ and posterior probabilities by running the algorithm independently 
for $K = 0$ to $K = \Kmax$, the total computation cost is $O(\Kmax^2/2)$.
 On the other hand, the proposed method requires a computation time of $O(K_0)$, 
 since the model generally searches the most probable model within the candidates. This is a significant speedup of $O(\Kmax/2)$, which may be beneficial 
under conditions where a large number of peaks is assumed in the data. 
However, as emphasized the above, sampling periods necessary for both methods
vary considerably. Under certain applications and depending on the size of the candidate model set,
 the proposed method may perform poorly compared with the conventional method.
  While not investigated in detail, these results encourage a hybrid application of the proposed model and the conventional method, where the proposed method is used to downsize the candidate model set to few probable models, and the conventional method is used on the remaining candidates for precise analysis.
\section{Simulation on Reflectance Spectra of Olivine Mixtures}
Encouraged by the performance on synthetic data, we applied the proposed method to the reflectance spectra of several mixtures of fayalite and forsterite, which constitute the mineral olivine.
Previous studies based on free-hand fitting and nonlinear fitting indicate that the reflectance spectrum of olivine shows three absorption bands at approximately $0.9, 1.1$, and $1.3$ \SI{}{\micro\meter}, with each band 
shifting its absorption position to higher wavelengths with decreasing forsterite mixture ratio\cite{burns1970, sunshine1998, David2013}.
 Mineralogical research indicate that the $0.9, 1.3$ \SI{}{\micro\meter} band originates from the M1 site in the olivine crystal, whereas the $1.1$ \SI{}{\micro\meter} band is attributed to the M2 site in the crystal \cite{burns1970}. 
 Since the absorption energy is inversely proportional to the ion ratio located in each site\cite{sunshine1998}, the absorption wavelength increases as the ions in each site are replaced from magnesium ions originating from the forsterite mixture 
 to larger iron ions originating from the fayalite mixture.
  The quantitative difference in the rate of decrease therefore depends on the coefficient, 
  which is assumed to be different for the two M1 and M2 cation sites. 
Correspondingly, the relative intensity of the M2 absorption is reported to increase with the forsterite content, whereas the two M1 bands (0.9 and 1.3 \SI{}{\micro\meter} absorption bands) conserve their intensity ratio\cite{sunshine1998}. 
Our research aims to confirm these properties from a purely data-driven perspective.
\\
\indent We describe the reflectance spectra using the modified Gaussian model with a continuum background modelled as a linear function of energy\cite{sunshine1990}. 
Given then wavelength $x$ and observed reflectance $y$, the model is given by 
\begin{equation}
    -\log y = c_0 + \frac{c_1}{x} + \sum_{k = 1}^K a_k \exp\qty[-\frac{\rho_k}{2}(x-\mu_k)^2] + \epsilon, \ \epsilon \sim \mathcal{N}(0, 1/b).
\end{equation}
The spectral data set used in this study was collected from the NASA/Keck Reflectance Experiment
Laboratory (RELAB) at Brown University and is shown in Table \ref{table:data}.
 Here, the Fo number is defined by the molar percentage of magnesium ions against the total number of magnesium and iron ions in the sample.
  The wavelength region used in the simulation for each sample was from 0.6 to 2.5 \SI{}{\micro\meter}, and the number of data points was $n = 381$. 
  \begin{figure*}[b]
      \centering
      \includegraphics{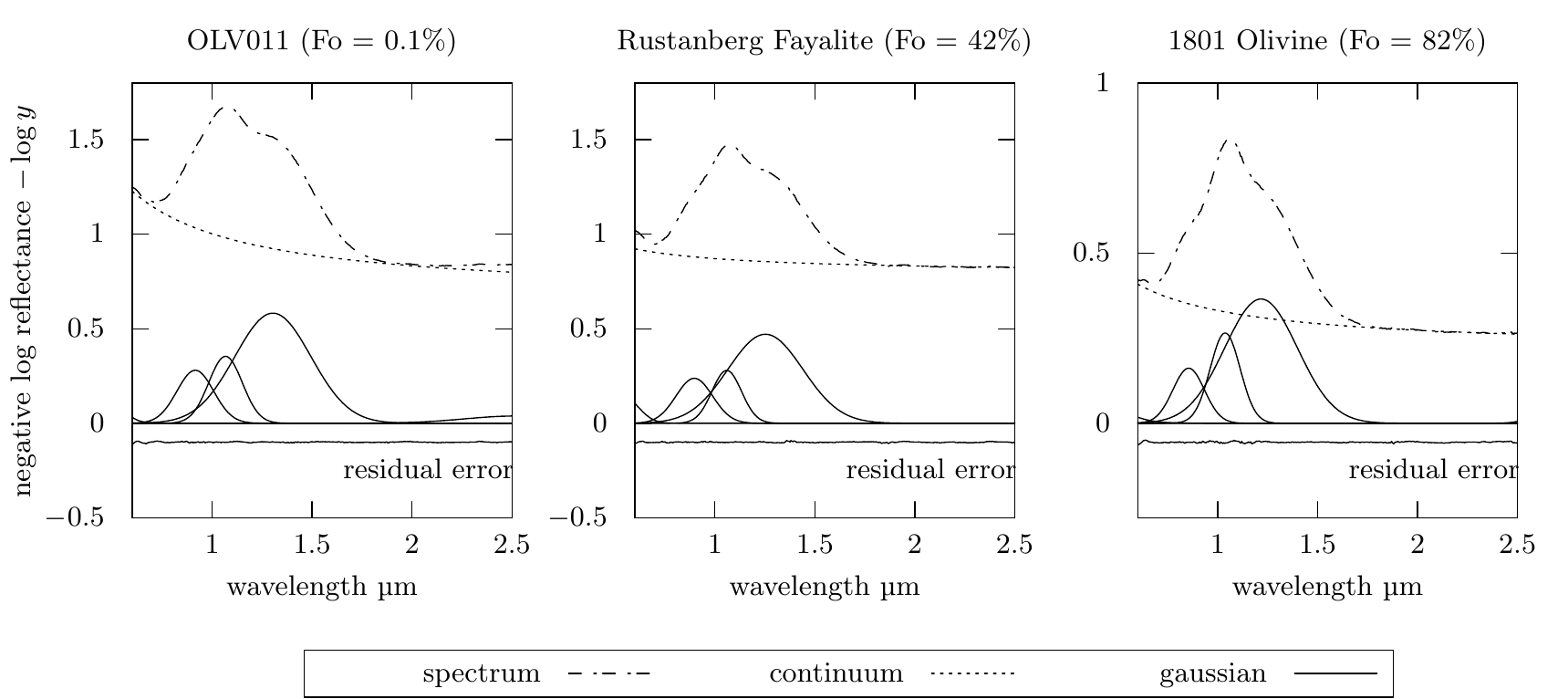}
      \caption{Deconvolution results for olivine samples OLV011, Rustanberg Fayalite, and 1801 Olivine. }
      \label{fig:data_deconv}
  \end{figure*}
  \begin{figure*}[t]
        \begin{minipage}[t]{.48\textwidth}
              \centering
              \includegraphics{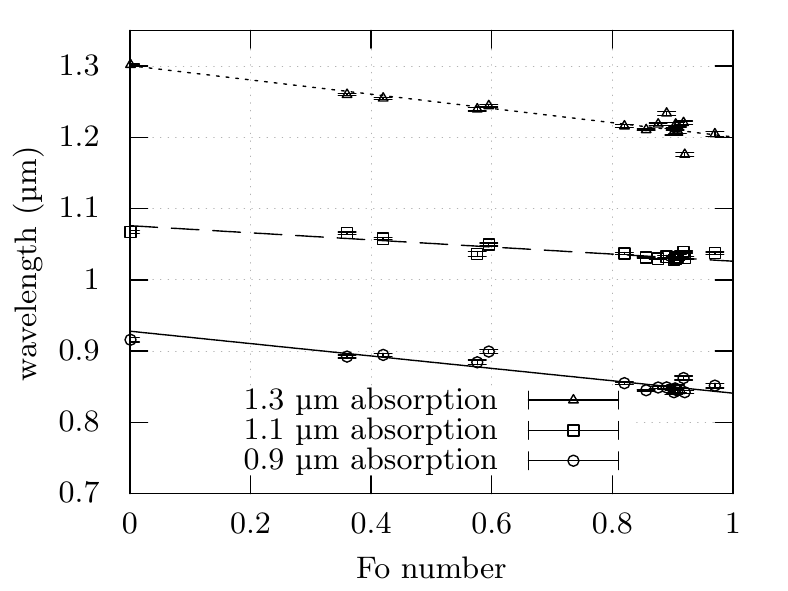}
              \caption{Center positions of the $0.9, 1.1,$ and $1.3$ \SI{}{\micro\meter} bands. The dotted, dashed, and solid lines have slopes of $-0.087\pm8$, $-0.049\pm4$, and $-0.100\pm5$ $\rm \mu m$ per mixture ratio, respectively, indicating a similarity among the M1 bands, and a dissimilarity between the M1 and M2 bands.}
       \label{fig:bandline}
       \end{minipage}
       \quad
        \begin{minipage}[t]{.48\textwidth}
               \centering
              \includegraphics{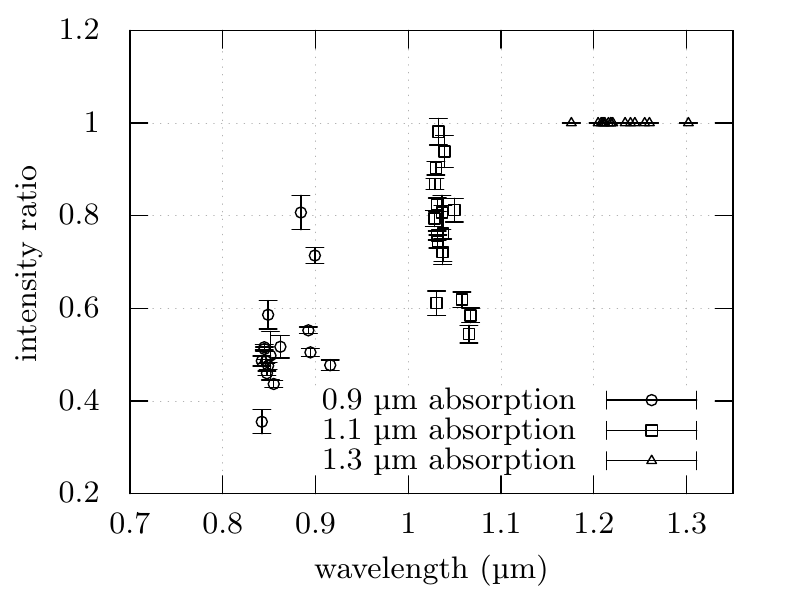}
              \caption{The intensity of the $0.9, 1.1$, and $1.3$ \SI{}{\micro\meter} bands normalized to the $1.3$ \SI{}{\micro\meter} band. The error bar for the center position was omitted for readability.}
              \label{fig:bandstrength}
        \end{minipage}\hfill%
  \end{figure*}
  \begin{figure}
      \centering
      \includegraphics{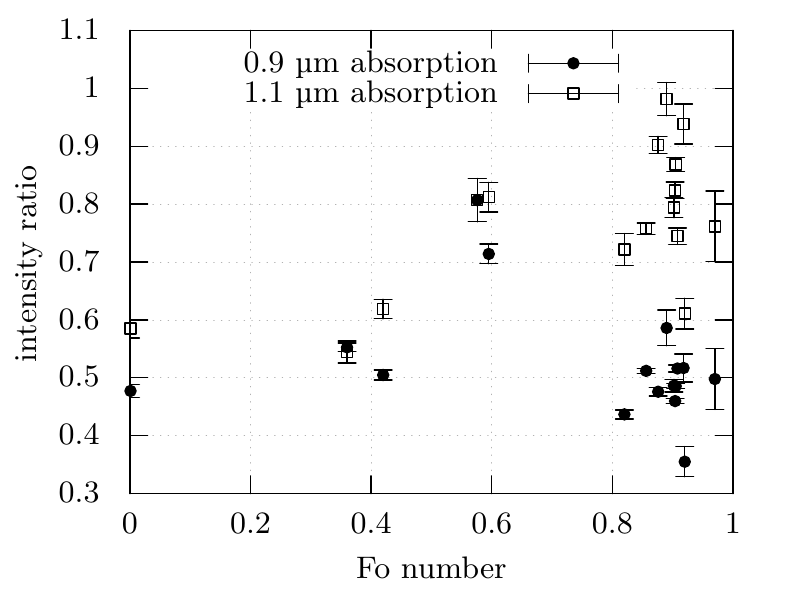}
      \caption{Intensity ratios of the $0.9$ and $1.1$ \SI{}{\micro\meter} bands (see Fig. \ref{fig:bandstrength}) against Fo number.}
      \label{fig:trend}
  \end{figure}
  \begin{table*}[h]
\centering
\caption{Fo numbers of olivine reflectance spectral data used in this study. The Fo numbers were obtained from Trang et al.\cite{David2013}.}
  \begin{tabular}{C{0.25\textwidth} | R{0.15\textwidth}}
    RELAB sample name & Fo number (\%) \\\hline
    OLV011 & 0.1 \\
    Franklin Fayalite & 36.0  \\
Rustanberg Fayalite & 42.0 \\
    OLV022 & 57.6 \\
     OLV020 &59.5 \\
    1801 Olivine & 82.0 \\
     Hawaii Olivine & 84.0 \\
    OLV002 & 85.6 \\
     OLV025 & 85.7\\
  \end{tabular}
  \begin{tabular}{C{0.25\textwidth} | R{0.15\textwidth}}
     RELAB sample name & Fo number (\%)  \\\hline
     OLV021 & 85.9 \\ 
    OLV201& 87.6  \\
    OLV005 & 90.2 \\
    OLV003 & 90.2  \\
    OLV013& 90.5  \\
     OLV102& 90.8 \\
     OLV012& 91.8 \\
    Apache Forsterite & 92.0 \\ 
    OLV007 & 96.9  
  \end{tabular}
  \label{table:data}
\end{table*}
The prior densities for the continuum background and peak parameters were set independently and defined as 
\begin{gather}
    p(\Theta_K, c_0, c_1) = p(c_0, c_1) \prod_{k = 1}^K \varphi(\theta_k),\\
    p(c_0,c_1) = \mathcal{N}(c_0 ; -\log y_{x=2.5\ \text{\SI{}{\micro\meter}} }, 0.1^2)\ {\rm Gam}(c_1;1, 10),\\
    p(\theta_k) = {\rm Gam}(a_k ; 3, 2)\ {\rm Gam}(\rho_k ; 5, 0.04)\ \mathcal{N}(\mu_k; 1.25, 0.4^2),
\end{gather}
where $y_{x = 2.5 \\text{\SI{}{\micro\meter}}}$ denotes the value of the reflectance data at $x = 2.5$ \SI{}{\micro\meter}.
 The prior density of $c_0$ is set to this particular value under the assumption that above wavelength $2.0$ \SI{}{\micro\meter}, the energy–linear 
 $c_1$ component of the continuum can be ignored, and relevant absorption features disappear, thus revealing the constant part of the continuum. 
 The prior density of $\mu_k$ reflects our prior knowledge of the positions of absorption bands. \\
\indent The number of replicas was set to $L  =72$, with the replica inverse temperatures set to \[
b_l = \begin{cases}
0 & (l = 1)\\
432000 \cdot1.2^{l - L}  & (\text{otherwise})
\end{cases}.
\]
The number of iterations was set to 100,000 MCSs for burnin and 50,000 MCSs for sampling.
 To determine the peak parameters, we take the empirical Bayesian approach; 
 only the data that correspond to model $(K^\star, b^\star)$ were accumulated, where \begin{equation}
b^\star = \argmin_{b\in \qty{b_l}_{l = 1}^L} F_n(b), \quad K^\star = \argmax_{K} p(K | \mathcal{D} , b = b^\star) .
\end{equation}
This approach allows us to deconvolute without prior knowledge on the noise level of the data; the scheme is purely data-driven. \\
\indent Deconvolution results on a subset of the olivine datum are given in Fig. \ref{fig:data_deconv}. 
Of the collected olivine samples, deconvolution results on OLV025 and OLV021 did not indicate the three particular bands. 
Excluding these two samples, the center positions of the M1 and M2 bands are given in Fig. \ref{fig:bandline}. 
As evident from the graph, the positions decline with respect to magnesium content, with the 0.9 and 1.3 \SI{}{\micro\meter} bands declining at
 slopes of $-0.087\pm8$ and $-0.100\pm5$ \SI{}{\micro\meter} per Fo number, whereas the $1.1$ \SI{}{\micro\meter} band declines at approximately half that rate of $-0.049\pm4$ \SI{}{\micro\meter} per Fo number. 
This is in agreement with results obtained by Sunshine and Pieters using nonlinear regression, who reported slopes of $-0.083, -0.045,$ and $-0.10$ \SI{}{\micro\meter} for the $0.9, 1.1,$ and $1.3$ \SI{}{\micro\meter} bands, respectively\cite{sunshine1998}.  \\
\indent The relative intensities of the absorption bands excluding the OLV025 and OLV021 samples is given in Fig. \ref{fig:bandstrength}. 
Excluding the outliers, the M1 bands have a relatively constant intensity ratio. 
As evident from Fig. \ref{fig:trend}, while the intensity ratio between the two M1 bands does not show a clear trend against magnesium content, 
the relative intensity of the M2 band shows a weak increasing trend against magnesium content, which is consistent with the results of previous research\cite{burns1970,sunshine1998}. 
Our results from the intensity ratio provide some evidence for coupling between the $0.9$ and $1.3$ \SI{}{\micro\meter} bands. \\
\indent Our results give information-based support to mineralogical knowledge obtained in previous research using nonlinear regression. 
On the other hand, our method was not successful in deriving the properties of fosterite/fayalite mixture spectra in some data sets. 
This indicates that our modelling of olivine spectra was inconsistent with the spectral data. 
A few of the reasons for inconsistency include an oversimplification of the continuum, 
as well as the lack of considering impurities and noise level dependence on wavelength or signal intensity. 
Nevertheless, these issues can be resolved at the data modelling level, and our method is expected to perform successfully posterior to these updates. 
\section{Conclusion}
In this study, we developed a deconvolution scheme designed to infer the number of spectral peaks and the optimal peak parameters for regression simultaneously. 
This new method is a combination of replica exchange MCMC and RJMCMC methods, 
with adaptive step-size tuning for the optimal proposal of MCMC moves in each replica.
Our method is effective when the number of peaks within spectral data is 
unknown prior to the knowledge that the simulation for each candidate model is computationally intensive. \\
\indent Compared with a sweep simulation over the candidate models, 
the proposed method must accumulate more samples to obtain the Bayesian free energy, and correspondingly a nonbiased population. 
On the other hand, the proposed method is capable of model selection within a 
considerably small number of MCSs. 
In the case where only a single optimal model among the candidates is of concern, one may exploit this advantage by using the proposed method for identifying this optimal model
 and using the conventional, single model MCMC sampler for a concentrated study. \\
\indent Nevertheless, the application of the proposed method to reflectance spectra of olivine with different mixture ratios of forsterite and fayalite indicated results similar to those obtained by 
geologists using nonlinear regression on the data sets. Our success suggests that the proposed method is practical and applicable to other deconvolution problems. 

\section*{Acknowledgments}
      The authors are grateful to Shiro Takagi for valuable discussions and for proofreading the manuscript.
      We also thank the anonymous reviewer for insightful comments significantly improving the manuscript.
       This work was supported by CREST (JPMJCR1761) from the Japan Science and Technology Agency (JST). 
\bibliographystyle{jpsj}
\bibliography{ref}

\begin{thebibliography}{10}

\bibitem{sunshine1990}
J.~M. Sunshine, C.~M. Pieters, and S.~F. Pratt: J. Geophys. Res. Solid Earth
  {\bfseries 95} (1990) 6955.

\bibitem{Nagata2012}
K.~Nagata, S.~Sugita, and M.~Okada: Neural Networks {\bfseries 28} (2012) 82.

\bibitem{ikeda2015}
S.~Ikeda and M.~Kotani: {\em A New Direction in Mathematics for Materials
  Science} (Springer, 2015), Vol.~1.

\bibitem{Bishop}
C.~M. Bishop: {\em Pattern Recognition and Machine Learning} (Springer, 2006).

\bibitem{Akaike1980}
H.~Akaike: Trabajos de Estadistica Y de Investigacion Operativa {\bfseries 31}
  (1980) 143.

\bibitem{watanabe2010}
S.~Watanabe: Neural Networks {\bfseries 23} (2010) 20 .

\bibitem{green1995}
P.~J. Green: Biometrika {\bfseries 82} (1995) 711.

\bibitem{Razul2001}
S.~{Gulam Razul}, W.~J. {Fitzgerald}, and C.~{Andrieu}: Nucl. Instrum. Methods.
  Phys. Res. B {\bfseries 497} (2003) 492.

\bibitem{Han2020}
N.~Han and R.~J. Ram: Comput. Stat. Data. Anal. {\bfseries 143} (2020) 106846.

\bibitem{hukushima1996}
K.~Hukushima and K.~Nemoto: J. Phys. Soc. Jpn. {\bfseries 65} (1996) 1604.

\bibitem{jasra2005}
A.~Jasra, C.~C. Holmes, and D.~A. Stephens: Stat. Sci. {\bfseries 20} (2005)
  50.

\bibitem{tokuda2017}
S.~Tokuda, K.~Nagata, and M.~Okada: J. Phys. Soc. Jpn. {\bfseries 86} (2017)
  024001.

\bibitem{ogata1990}
Y.~Ogata: Ann. Inst. Stat. Math. {\bfseries 42} (1990) 403.

\bibitem{gelman1998}
A.~Gelman and X.-L. Meng: Stat. Sci. {\bfseries 13} (1998) 163.

\bibitem{tokuda2014}
S.~Tokuda, K.~Nagata, and M.~Okada: IPSJ Online Trans. {\bfseries 7} (2014) 20.

\bibitem{Katzgraber2019}
I.~Rozada, M.~Aramon, J.~Machta, and H.~G. Katzgraber: Phys. Rev. E {\bfseries
  100} (2019) 043311.

\bibitem{Swensden1989}
A.~M. Ferrenberg and R.~H. Swendsen: Phys. Rev. Lett. {\bfseries 63} (1989)
  1195.

\bibitem{garthwaite2016}
P.~H. Garthwaite, Y.~Fan, and S.~A. Sisson: Commun. Stat. - Theory Methods
  {\bfseries 45} (2016) 5098.

\bibitem{robbins1951}
H.~Robbins and S.~Monro: Ann. Math. Stat. {\bfseries 22} (1951) 400.

\bibitem{burns1970}
R.~G. Burns: Am. Mineral. {\bfseries 55} (1970) 1608.

\bibitem{sunshine1998}
J.~M. Sunshine and C.~M. Pieters: J. Geophys. Res. Planets {\bfseries 103}
  (1998) 13675.

\bibitem{David2013}
D.~Trang, P.~G. Lucey, J.~J. Gillis-Davis, J.~T.~S. Cahill, R.~L. Klima, and
  P.~J. Isaacson: J. Geophys. Res. Planets {\bfseries 118} (2013) 708.

\end{thebibliography}
\end{document}